\documentclass[pra, showpacs, twocolumn, floatfix,superscriptaddress]{revtex4}
\usepackage{graphicx}
\usepackage{epstopdf}
\usepackage{amsmath, amsfonts, amssymb, bm, ulem, color}
\begin{document}
\title{Fast phonon dynamics of a nanomechanical oscillator due to cooperative effects}
\author{Victor \surname{Ceban}}
\affiliation{Max-Planck-Institut f\"{u}r Kernphysik, Saupfercheckweg 1, D-69117 Heidelberg, Germany }
\affiliation{Institute of Applied Physics, Academy of Sciences of Moldova, 
Academiei str. 5, MD-2028 Chi\c{s}in\u{a}u, Moldova}
\author{Paolo \surname{Longo}}
\affiliation{Max-Planck-Institut f\"{u}r Kernphysik, Saupfercheckweg 1, D-69117 Heidelberg, Germany }
\author{Mihai A. \surname{Macovei}}
\affiliation{Max-Planck-Institut f\"{u}r Kernphysik, Saupfercheckweg 1, D-69117 Heidelberg, Germany }
\affiliation{Institute of Applied Physics, Academy of Sciences of Moldova, 
Academiei str. 5, MD-2028 Chi\c{s}in\u{a}u, Moldova}
\date{\today}
\begin{abstract}
We investigate the coupled-system dynamics of two-level quantum dots placed on a vibrating nanomechanical resonator.
The ensemble of quantum dots exhibits superradiance features which are transferred to the 
mechanical degrees of freedom representing fast quantum dynamics and enhanced phonon emission in a nanomechanical setup, 
resembling of the superradiance effect.
\end{abstract}
\maketitle
\section{Introduction}
First discovered by Dicke \cite{dic}, the phenomenon of superradiance (SR) plays an important role in the context of collective quantum processes. 
SR can be understood as a collective giant dipole resulting from an ensemble of closely packed emitters. As a result, the spontaneous emission dynamics 
is modified, leading to a reduced radiative lifetime as well as an enhanced emission intensity (proportional to the square of the number of excited emitters). 
Interestingly, the speed up of the spontaneous decay also occurs for single-photon excitations \cite{scu1}, {\it i.e.}, when only a single emitter of the 
atomic sample is excited. From early experiments with driven gases \cite{skr}, SR has been investigated intensely in various theoretical and experimental 
approaches \cite{gro, kei, andr, scu2, con}. Since then, remarkable results have been achieved and SR was observed in molecular aggregates \cite{boe} 
and crystals \cite{fro}, as well as in Bose-Einstein condensates where Dicke phase transition occurs \cite{bau}. Another important goal was achieved in 
the realm of condensed matter physics, where SR was observed within a collection of quantum dots (QDs) \cite{sch}. Moreover, single-photon SR has 
been reported recently for artificial atomic samples \cite{tig}, whereas an analytical framework for the engineering of single-photon SR in extended 
media was developed in \cite{lke}. Quantum wells allow for even more exotic SR behavior as their excited states---the excitons---are created in a 
two-dimensional layer \cite{con}, which is different from the QD case where excitons are well localized in space.

In condensed matter systems, it is also possible to couple quantum optical systems to quantum mechanical vibrations at mesoscopic scales.
Nowadays, such scenarios are investigated in the field of optomechanics \cite{fav}. Advances in the fabrication techniques of 
optomechanical devices \cite{asp} have led to a series of fundamental experimental achievements like near-ground state quantum 
cooling \cite{oco, roc}, the phonon laser analogue \cite{vah}, or squeezing of the mechanical motion \cite{wol}. 
Furthermore, the manipulation and confinement of phonons may be facilitated or enhanced when quantum mechanical resonators 
are considered. Therefore, single-mode phonon fields may be obtained through different architectures of resonators such as 
nanobeams \cite{roc}, vibrating membranes \cite{tho, bag}, or multilayered acoustic nanocavities \cite{soy}.

Motivated by these achievements, we consider here a setup where collective radiative effects and nanomechanical motion are brought together.  
In particular, we envision a model of a collection of initially excited two-level QDs which are embedded on a nanomechanical resonator. The QDs 
are spatially arranged to allow for a superradiant collective decay. The nanomechanical vibrations couple to the QDs excited states, leading to a 
modified phonon dynamics. In fact, we observe that the resulting dynamics has cooperative features in both the phonon emission time scales 
and the intensity of the generated phonon field. In other words, we investigate phonon superradiance in a nanomechanical setup. 
This effect may improve the optical mass sensing or biosensing schemes described in \cite{k_zhu,ufn}, respectively, or to 
enhance ultra-weak signal detections \cite{ufn,brag}. Notice that the phenomenon of phonon superradiance was investigated in a different context 
in \cite{supf,subf}, for instance.

This paper is organized as follows. In Sec. II, we describe the analytic model and analyze the relevant equations of motion via the system's master 
equation. In Sec. III, we solve these equations and discuss the resulting dynamics of the QDs' excitation  and the phonon field. A summary is given 
in Sec. IV.

\section{The Model and Equations of Motion}
\begin{figure}[t]
\centering
 \includegraphics[width= 8 cm ]{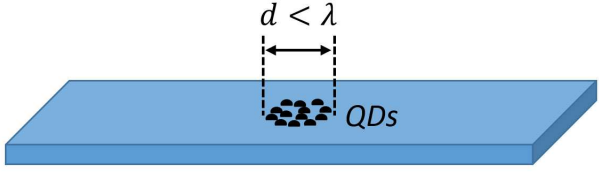}
\caption{\label{fig0} 
(color online) {Schematic of the model: An initially excited ensemble of two-level quantum dots (QDs) are fixed on a 
vibrating nanomechanical resonator. The linear dimension $d$ of the sample is smaller than the relevant transition 
photon wavelength $\lambda$, {\it i.e.,} $d < \lambda$. However, this condition can be relaxed for certain sample's 
geometries \cite{gro}.}}
\end{figure}

We consider a collection of $N$ initially excited two-level quantum dots (QDs) fixed on a vibrating membrane (see Fig. \ref{fig0}). The membrane acts as 
a quantum harmonic oscillator (frequency $\omega$) and is coupled to an environmental thermal reservoir of temperature $T$. The mechanical oscillator 
couples to each QD of the sample with a coupling constant $\eta$. At lower environmental temperatures, one can consider the fundamental mechanical 
mode exclusively since other modes at higher frequencies contribute only weakly to the whole quantum dynamics. This is the 
case if the length $L$  of the nanomechanical resonator is considerably bigger than its width $l$ and thickness $a$ \cite{wil}. 
For $L \sim 10^{3}{\rm nm}$, $a \sim 30 {\rm nm}$ and $l \sim 100{\rm nm}$ one can still have a sufficient number of quantum dots fixed on 
the nanomechanical resonator in order to the superradiance effect to occur (the sizes of quantum dots are approximately within few to several 
nanometers). The corresponding system Hamiltonian is
\begin{eqnarray}
H = \hbar\omega b^{\dagger}b + \hbar \omega_{\mathrm{qd}}S_{z} + \hbar \eta S_{22} ( b + b^{\dagger}).
\label{Htot}
\end{eqnarray}
Here, the first term is the membrane's free single-mode Hamiltonian, expressed via the phonon annihilation and creation operators 
$b$ and $b^{\dagger}$, respectively, that obey the usual bosonic commutation relations $[b,b^{\dagger}]=1$ and 
$[b,b]=[b^{\dagger},b^{\dagger}]=0$. The second term represents the free Hamiltonian of the QD sample, where 
$\omega_{\mathrm{qd}}$ denotes the QD's transition frequency. In the second term,
the collective inversion operator of the QD sample reads $S_{z} = \sum _{j=1}^{N} {S_{z}^{(j)}}$. 
Here, the $j{\rm th}$ QD is described by its excited $\vert e \rangle _{j}$ and ground $\vert g \rangle _{j}$ levels.
The corresponding single QD operators are
$S^{+}_{j} = \vert e \rangle _{j} {}_{j} \langle g \vert $, 
$S^{-}_{j} = \vert g \rangle _{j} {}_{j} \langle e \vert$, and 
$S_{z}^{(j)} = \bigl( \vert e \rangle _{j} {}_{j} \langle e \vert  - \vert g \rangle _{j} {}_{j} \langle g \vert \bigr )/2$. 
Finally, the last term in Hamiltonian 
(\ref{Htot}) represents the interaction of the QD sample with the phonon field. In particular, we have considered that the membrane's 
spatial scale is larger than the extent of the QD sample (see Fig.~\ref{fig0}). Consequently, the coupling strengths of each QD with 
the vibrational degrees of freedom are identical and have the same magnitude $\eta$. The collective operator for the QDs' upper state 
is defined as $S_{22} = \sum _{j=1}^{N} {\vert e \rangle _{j} {}_{j} \langle e \vert }$. 

The system dynamics is described by the master equation for the density matrix operator $\rho$ as \cite{martin}
\begin{eqnarray}
\frac{\partial \rho}{\partial t} &=& - \frac{i}{\hbar} [ H, \rho ]  + \kappa \bar{n} \mathcal{L} (b^{\dagger}) 
+ \kappa ( 1+ \bar{n} ) \mathcal{L} (b) \nonumber \\
&+& \gamma \mathcal{L} (S^{-}).
\label{Meq}
\end{eqnarray}
In this master equation, the damping terms are expressed by the the Liouville superoperator $\mathcal{L}$, which,
for a given operator $\mathcal{O}$, is defined as
$\mathcal{L} (\mathcal{O}) = 2 \mathcal{O} \rho \mathcal{O}^{\dagger} - \mathcal{O}^{\dagger} \mathcal{O} \rho - \rho \mathcal{O}^{\dagger} 
\mathcal{O}$. 
Here, as usually, the first term of the master equation represents the coherent evolution as determined through Hamiltonian~(\ref{Htot}),
whereas the second and the third terms, respectively, denote the pumping and the damping of the mechanical resonator via the 
environmental thermal reservoir. The reservoir is described by the mean phonon number $\bar n =1/\bigl [\exp{(\hbar\omega/k_{B}T)}-1 \bigr ]$
and  the damping rate~$\kappa$. Here, $k_{B}$ is the Boltzmann constant and $T$ denotes the reservoir's temperature.
The last term characterizes the collective spontaneous emission from the sample of QDs as a result 
from the interaction of the sample with the environmental electromagnetic vacuum field modes. 
This term is characterized by the single-QD spontaneous decay rate $\gamma$, 
where we have assumed that the spatial separation between different QDs in the sample
is much smaller than the QDs' transition wavelength, {\it i.e.}, we assume a small-volume sample.
Hence, the collective operator for the QDs in the master equation (\ref{Meq}) is simply the sum of single 
QD operators, {\it i.e.}, $S^{\pm} = \sum _{j=1}^{N} {S^{\pm}_{j}}$. 

The equations of motion for  the mean phonon number $\langle b^{\dagger}b \rangle$, the population inversion $\langle S_{z} \rangle$,
and the corresponding correlators follow from master equation (\ref{Meq}), resulting in 
\begin{eqnarray}
\label{eq:eqom}
\frac{\partial \langle b^{\dagger}b \rangle}{\partial t} &=& i \eta \lbrace \langle S_{z} b \rangle - 
\langle S_{z} b^{\dagger} \rangle +j \langle b \rangle - j \langle b^{\dagger} \rangle \rbrace  \nonumber \\ 
&-& 2 \kappa \langle b^{\dagger}b \rangle + 2 \kappa \bar{n} , \nonumber \\ 
\frac{\partial \langle S_{z} b \rangle}{\partial t} &=& - (\kappa + 2 \gamma + i \omega) \langle S_{z} b \rangle 
+ 2 \gamma \langle S_{z}^{2} b \rangle  \nonumber \\ 
&-&i \eta \lbrace \langle S_{z}^{2} \rangle + j\langle S_{z} \rangle \rbrace - 2 \gamma j (j+1) \langle b \rangle , \nonumber \\
\frac{\partial \langle S_{z} b^{\dagger} \rangle}{\partial t} &=& - (\kappa + 2 \gamma - i \omega) \langle S_{z} b^{\dagger} \rangle 
+ 2 \gamma \langle S_{z}^{2} b^{\dagger} \rangle \nonumber \\ 
&+&i \eta \lbrace \langle S_{z}^{2} \rangle + j\langle S_{z} \rangle \rbrace - 2 \gamma j (j+1) \langle b^{\dagger} \rangle  , \nonumber \\
\frac{\partial \langle b \rangle}{\partial t} &=& -(\kappa+i \omega) \langle b \rangle - i \eta \lbrace \langle S_{z} \rangle +j \rbrace , \nonumber \\
\frac{\partial \langle b^{\dagger} \rangle}{\partial t} &=& -(\kappa -i \omega) \langle b^{\dagger} \rangle 
+ i \eta \lbrace \langle S_{z} \rangle +j \rbrace , \nonumber \\
\frac{\partial \langle S_{z} \rangle}{\partial t} &=& - 2 \gamma \lbrace \langle S_{z} \rangle - \langle S_{z}^{2} \rangle + j(j+1) \rbrace .
\label{syst}
\end{eqnarray}
In deriving these equations, 
we have used $j \equiv N/2$ and the relations $S_{z} = S_{22}-j$ and $S_{z}^{2} + (S^{+} S^{-} + S^{-} S^{+})/2 =j(j+1)$.
Note that the collective operators satisfy $[S^{+},S^{-}]= 2 S_{z}$ and $ [S_{z},S^{\pm}]= \pm S^{\pm} $.
In the next Section, we discuss how we close the set of equations~({\ref{eq:eqom}}).

\section{Results and Discussions}
In the following, we present 
two methods of solving the system of equations (\ref{syst}). 
It is possible to close the system when the number of 
elements of the sample is either small or big. 

\subsection{Small Number of QDs}
For only a few QDs, it is possible to solve the system by using the explicit expression of the $S_{z}^{2}$ depending terms, {\it i.e.}, by 
considering the equations of motion of every $S_{z}^{(i)} S_{z}^{(j)}$ term. Here, we will consider the simplest cases of a single QD, 
{\it i.e.}, $N=1$ and a pair of collectively interacting quantum dots (see Appendix A) placed on the membrane. In the first case, $N=1$,
$S_{z}^{2} = 1 /4$ and, hence, the system is closed and analytically solvable. Considering that the QD is initially excited while the membrane 
is in thermal equilibrium with the reservoir, the temporal behavior of the mean phonon number is given by
\begin{eqnarray}
\langle b^{\dagger}b \rangle &=& \bar{n} + \bar{a} e^{-2 \gamma t} -(\bar{a} + \bar{b} \bar{c}) e^{-2 \kappa t} \nonumber \\
&+& \bar{b} e^{-(2 \gamma + \kappa) t} \left( \bar{c} \cos{\omega t} - \bar{d} \sin{\omega t} \right) ,
\label{oneAtom}
\end{eqnarray}
where $\bar{a} = \eta^{2} / \bigl[(\kappa - \omega )(\kappa ^{2} + \omega ^{2})\bigr]$, 
$\bar{b} = 2 \eta^{2} / (\bar{c}^{2}+\bar{d}^{2})$, 
$\bar{c} = 2 \gamma \kappa - \kappa ^{2} - \omega ^{2}$, 
and $\bar{d} = 2 \omega \kappa$. 
Likewise, the 
temporal evolution of the 
QD population inversion reads \cite{martin}
\begin{figure}[t]
\centering
\includegraphics[width= 8 cm ]{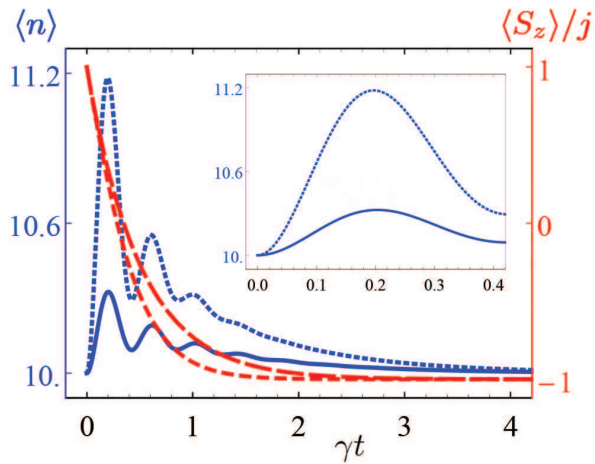}
\caption{\label{oneQD} 
(color online) Temporal evolution of the mechanical resonator's mean phonon number $ \langle n \rangle$ 
and the population inversion $ \langle S_{z} \rangle$ for a single-QD sample (solid and long-dashed lines, 
respectively) as well as for $N=2$ (dotted and short-dashed curves, respectively). For this plot, the 
parameters are: $\omega / \gamma= 15$, $\eta/\gamma=5$, $\kappa/\gamma=0.5$, and $\bar{n}=10$. 
The inset shows the mean phonon number for $N=1$ and $N=2$ at the beginning of the time-evolution.}
\end{figure}
\begin{eqnarray}
\langle S_{z}(t) \rangle = -1/2 + e^{-2 \gamma t }.
\label{szz}
\end{eqnarray}
For initially excited $N=2$ collectively interacting qubits, we solved numerically the system of equations which are given in the Appendix A. 
The initial condition for the involved variables are: $\langle S_{z}\rangle =1$, $\langle S^{+}S^{-}\rangle =2$ and 
$\langle b^{\dagger}b\rangle =\bar n$, whereas all other variables are zero at $t=0$.

In Figure~\ref{oneQD}, we depict the dynamics of $\langle b^{\dagger}b \rangle$ and $\langle S_z \rangle$ for the case when a single QD 
or a pair of collectively interacting QDs are placed on the membrane. We see that the mean phonon number first slightly increases and 
oscillates as a result of the interaction between the QD sample and the mechanical vibrations. However, for $N=2$ the mean vibrational 
phonon number is enhanced in comparison to single-qubit case although the dynamics is not faster (i.e., one needs larger ensembles to 
see a rapid evolution) in spite of the fact that the inversion decays faster for $N=2$.  After the QD's decay, the membrane returns 
back to thermal equilibrium since in Hamiltonian~(\ref{Htot}) phonons only couple to the QDs' excited states. Consequently, once the QD 
reaches its ground state, the phonon field is subject to damping due to the thermal reservoir. Another intrinsic propriety of the model is that 
the behavior of the QD population decay and its fluorescence dynamics are not affected by the phonon activity in this model. This propriety 
is observed in the analytic expression of $\langle S_{z} \rangle$ (see also Appendix A) and it is also valid for the additional results obtained 
later for a more numerous QD collection. This is the case if the preparation time, $\Delta t$, of the initial excited state is fast, {\it i.e.}, it 
takes place on a time-scale shorter than $\eta^{-1}$.
\subsection{Large QD Samples}
For samples of $N\gg 1$ QDs, the system of equations (\ref{syst}) can be closed by a factorization
of the correlations $\langle S_{z}^{2} \rangle$, $ \langle S_{z}^{2} b^{\dagger} \rangle$, 
and  $\langle S_{z}^{2} b \rangle$ according to Ref. \cite{andr}. 
\begin{figure}[t]
\centering
\includegraphics[width= 8cm ]{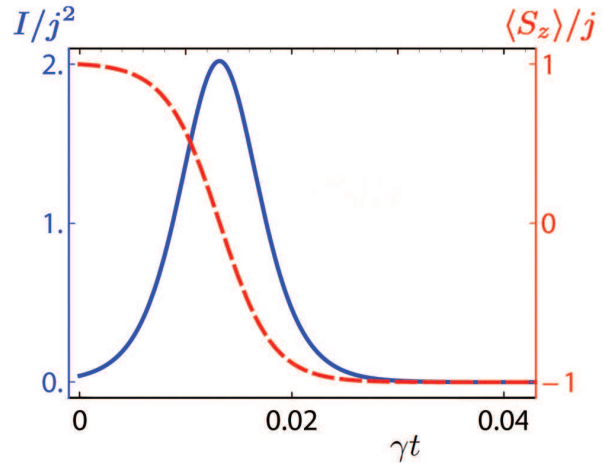}
\caption{\label{superFL} 
(color online) Temporal evolution of the collective population inversion 
$\langle S_{z}(t)\rangle /j$ for a collection of $N=200$ initially excited 
QDs (dashed curve) and their collective fluorescence intensity $I / j^{2} $ (solid line).
Note that $I(t) \propto -  \partial \langle S_{z} \rangle /\partial t $. }
\end{figure}

We start with  $\langle S_{z}^{2} \rangle \simeq\langle S_{z} \rangle ^{2}$, which
is equivalent to neglecting the fluctuations 
of the collective population inversion of the QD sample (and is reasonable as long as $N \gg 1$).
This assumption does not break the symmetry of the system as there 
are no new variables introduced.
In this context, the collective inversion operator $\langle S_{z}(t)\rangle$ 
simply becomes \cite{andr}
\begin{eqnarray}
\langle S_{z}(t)\rangle = -\frac{N}{2}\tanh\bigl[\frac{1}{2\tau_{R}}(t-t_{0})\bigr], 
\label{szc}
\end{eqnarray}
where $\tau_{R}=1/(2\gamma N)$ and $t_{0}=\tau_{R}\log(N)$. 
For the higher order correlations, the system symmetry is also maintained 
and we can choose either 
$\langle S_{z}^{2} b \rangle \simeq\langle S_{z} \rangle \langle S_{z} b \rangle$ 
or $\langle S_{z}^{2} b \rangle \simeq\langle S_{z} \rangle ^{2} \langle b \rangle$ 
for the decoupling, leading to the same results.
Similarly, we proceed with the $\langle S_{z}^{2} b^{\dagger} \rangle$ term.
Note that this scheme is justified for larger QD ensembles as well as phonon numbers.

Using the numerical solution of the system of equations (\ref{syst}), we depict the dynamics of the collective population inversion of the QD 
sample and the corresponding fluorescence intensity $I(t) \propto -  \partial \langle S_{z} \rangle /\partial t $
in Fig. \ref{superFL} for which
we have chosen $\langle S_{z} \rangle_{t=0}=j$, 
$\langle S_{z} b\rangle_{t=0}=\langle S_{z} b^{\dagger} \rangle_{t=0}=\langle b \rangle_{t=0}=\langle  b^{\dagger} \rangle_{t=0}=0$, and 
$\langle  b^{\dagger} b \rangle_{t=0} = \bar{n} $ as initial conditions.
Note that the two-level sample can be excited initially with a short laser pulse of duration $\Delta t < 1/ \eta$ 
in order to avoid the influence of vibrational phonons on the preparation stage. 
Similar to the single-QD case, the vibrations of the membrane do not affect the 
behavior of $\langle S_{z}(t)\rangle$ and $ \partial \langle S_{z}(t)\rangle /\partial t$, 
resembling the classical 
superradiance effect of a small volume sample.
Indeed, one obtains a decrease of the QDs lifetime which is inversely proportional to $N$, whereas the intensity increases 
proportional to $N^{2}$. 
Evidently, the maximum fluorescence intensity is reached for an equal number of QDs in the excited and ground states, 
respectively (see Figure~\ref{superFL}).
\begin{figure}[t]
\centering
\includegraphics[width= 8cm ]{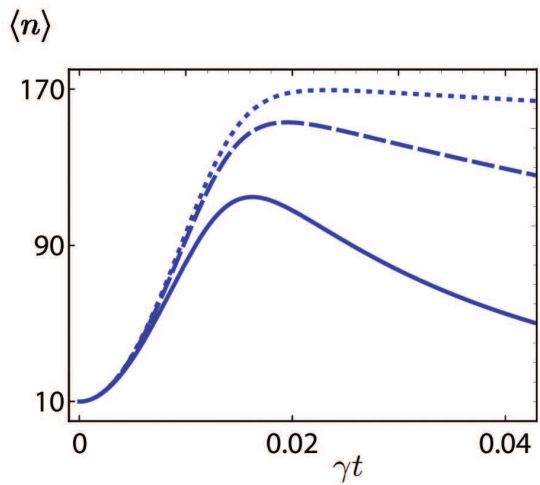}
\caption{\label{phon} 
(color online) Temporal evolution of the mechanical resonator's mean phonon number $\langle n \rangle$.
Here, a collection of $N=200$ QDs is excited initially.
The results are shown for different damping rates $\kappa$, {\it i.e.},
for $\kappa/\gamma = 1$ (dotted curve), $\kappa/\gamma = 5$ (dashed curve), and $\kappa/\gamma = 20$ (solid curve). 
For this plot, we further have
$\omega / \gamma= 50$, $\eta/\gamma=5$, and $\bar{n}=10$. }
\end{figure}

On the other hand, the phonon dynamics is affected by the collective effect within the QD sample as shown in Fig.~\ref{phon}. 
Here, a collective behavior is observed in both a reduced lifetime and an enhanced mean phonon number.
This can be explained as follows. The membrane's vibrations interact with the QDs' excited states and, therefore, 
the time scale for when phonons are created is related to the decay rate of the atomic sample. 
Then---in analogy to the single-QD case---when the QD sample approaches its collective ground state, 
the phonon dynamics becomes dominated by damping phenomena due to the environmental reservoir.
Hence, the mean phonon number decreases to its initial value $\bar{n}$, characterizing equilibrium with the thermal reservoir. 
Further, the superradiance effect exhibits a bell-like behavior for the collective intensity. One can identify this behavior in 
Fig.~\ref{phon} together with faster dynamics in comparison to the single-qubit case. However, the intensity does not simply 
scale as $N^{2}$ although one has a clear enhancement of phonon emission. In the absence of collective effects, we would have 
a similar behavior as shown in Fig.~\ref{oneQD} for $N=1$, i.e. no fast dynamics. However, the phonon number will be enhanced 
as well as a result of the coupling of many independent qubits to phonons.

Note that although the phonon cooperative effect is interconnected to the collective phenomena within the QD sample---and therefore 
to the number $N$ of QDs---the maximum mean phonon number is also determined by the damping rate $\kappa$. This can be seen in 
Fig.~\ref{phon} as the superradiant phonon emission increases in width and maximum for weaker damping ({\it i.e.} for smaller 
$\kappa$), resembling a good or bad cavity limit, respectively.

\section{Summary}
In conclusion, we have investigated the quantum dynamics of a coupled system composed of an ensemble of two-level quantum dots
that are fixed on a vibrating, nano-mechanical membrane. We have discussed the temporal evolution and the underlying equations of 
motion in detail. As the main result, we found that the QD sample exhibits superradiance features which are transferred to the vibrational 
degrees of freedom of the nanomechanical resonator, leading to phonon effects resembling superradiance in a nanomechanical setup. 
Furthermore, the detection of photon superradiance ensures the existence of phonon superradiance. Thus, our scheme may serve as a 
vibrational phonon detector in case of superradiance effects.

\section*{Acknowledgement}
We appreciate the helpful discussions with J\"{o}rg Evers and Christoph H. Keitel. P. L. is grateful for the hospitality of the 
Institute of Applied Physics of the Academy of Sciences of Moldova.  Furthermore, we acknowledge the financial support by 
the German Federal Ministry of Education and Research, grant No. 01DK13015, and Academy of Sciences of Moldova, 
grants No. 13.820.05.07/GF and 15.817.02.09F. 

\appendix
\section{N=2 collectively interacting QDs}
Here we present the equations of motion for a pair of collectively interacting QDs, $q1$ and $q2$, i.e.,
\begin{eqnarray}
\frac{\partial \langle S_{z} \rangle}{\partial t} &=& - 2 \gamma  \langle S^{+}S^{-} \rangle  ,
\nonumber \\
\frac{\partial \langle S^{+}S^{-} \rangle}{\partial t} &=& 8 \gamma \lbrace 1+ \langle S_{z} \rangle - \langle S^{+}S^{-} \rangle  \rbrace , 
\nonumber \\
\frac{\partial \langle b^{\dagger}b \rangle}{\partial t} &=& i \eta \lbrace \langle S_{z} b \rangle - 
\langle S_{z} b^{\dagger} \rangle + \langle b \rangle - \langle b^{\dagger} \rangle \rbrace  \nonumber \\ 
&-& 2 \kappa \langle b^{\dagger}b \rangle + 2 \kappa \bar{n} , 
\nonumber \\ 
\frac{\partial \langle S_{z} b \rangle}{\partial t} &=& - (\kappa  + i \omega) \langle S_{z} b \rangle 
- 2 \gamma \langle S^{+}S^{-} b \rangle  \nonumber \\ 
&-&i \eta \lbrace 2+2 \langle S_{z} \rangle - \langle S^{+}S^{-}  \rangle  \rbrace , \nonumber \\
\frac{\partial \langle S^{+}S^{-} b \rangle}{\partial t} &=& -(\kappa + 8 \gamma + i \omega) \langle S^{+}S^{-} b \rangle   \nonumber \\ 
&-& 2 i \eta \lbrace 1+ \langle S_{z} \rangle \rbrace + 8 \gamma \lbrace \langle b \rangle + \langle S_{z} b \rangle \rbrace, 
\nonumber \\ 
\frac{\partial \langle b \rangle}{\partial t} &=& -(\kappa+i \omega) \langle b \rangle - i \eta \lbrace 1+ \langle S_{z} \rangle \rbrace,
\label{syst2q}
\end{eqnarray}
where $S_{z}=S^{(q1)}_{z} + S^{(q2)}_{z}$ and $S^{\pm} = S^{\pm}_{q1} + S^{\pm}_{q2}$. The missing equations of motion can be obtained 
via hermitian conjugation of the last three equations in (\ref{syst2q}). This system of equations is solved numerically and the results are shown in 
Fig.~\ref{oneQD}. One can observe that the  vibrational mean phonon number is increased in comparison to the single-qubit case.


\end{document}